
\input phyzzx


\catcode`\@=11
\def\binrel@#1{\setbox\z@\hbox{\thinmuskip0mu
 \medmuskip\m@ne mu\thickmuskip\@ne mu$#1\m@th$}%
 \setbox\@ne\hbox{\thinmuskip0mu\medmuskip\m@ne mu\thickmuskip
 \@ne mu${}#1{}\m@th$}%
 \setbox\tw@\hbox{\hskip\wd\@ne\hskip-\wd\z@}}
\def\underset#1\to#2{\binrel@{#2}\ifdim\wd\tw@<\z@
 \mathbin{\mathop{\kern\z@#2}\limits_{#1}}\else\ifdim\wd\tw@>\z@
 \mathrel{\mathop{\kern\z@#2}\limits_{#1}}\else
 {\mathop{\kern\z@#2}\limits_{#1}}{}\fi\fi}
\def\utilde#1{\underset{\widetilde{\ }} \to #1}

\pubnum={206/COSMO$-$24}

\titlepage
\vskip 1cm
\title{\bf{New Vector Field and BRST Charges \nextline
       in 2-form Einstein Gravity}}
\vskip 1cm
\author{Akika Nakamichi
        \footnote{\star}{e-mail address: akika@phys.titech.ac.jp}\
 and \  Tatsuya Ueno
        \footnote{\dagger}{e-mail address: tatsuya@phys.titech.ac.jp}  }
\vskip 1cm
\address{Department of Physics, Tokyo Institute of Technology
         \nextline Oh-okayama, Meguro-ku, Tokyo 152, Japan     }
\vskip 1 cm

\abstract{A new vector field is introduced into 2-form Einstein gravity
in four dimensions to restore a large symmetry of its topological
version.
 Two different expressions for the BRST charge are given in the system:
one of them associated with a set of irreducible symmetries and
the other with a set of on-shell reducible symmetries. }

\vfill
\eject

\sequentialequations

\baselineskip=24pt



 Several approaches have been developed in recent studies of quantum
gravity to clarify its various aspects.
 Among them, an approach based on a class of topological quantum field
theories, called topological gravity, is expected to elucidate the
global (topological) aspects of gravity.
 In Ref.[1], Lee and the present authors proposed a topological version
of four-dimensional Einstein gravity.
 This topological version is obtained by modifying an alternative
formulation of gravity enlightened by Capovilla et al.[2], in which
anti-self-dual 2-forms are used as fundamental variables, instead of the
metric.
 This formulation, which we call 2-form Einstein gravity, naturally
leads to the canonical formalism discovered by Ashtekar [3].
 Investigating the relation between 2-form Einstein gravity and
its topological version, we found that, in the presence of a
cosmological term, a unique quantum state in the latter turns out to be
one of the physical states in the former, that is, one of the exact
solutions of the Wheeler-DeWitt equation. \par

 The topological version is promoted back to Einstein gravity by adding
a certain term in the action and, due to the term, a large topological
symmetry is partially broken leaving diffeomorphism and local-Lorentz
symmetries.
 The situation is similar to the one between the massless and massive
U(1) gauge theories.
 As is well known, the gauge invariance of the former is explicitly
broken by the mass term.
 We then encounter singularities in the zero-mass limit because there is
no gauge prescription in the limiting action.
 To avoid the singularities, one needs the Stueckelberg field which
restores the broken gauge invariance in the massive theory [4]. \par

 In this paper, we introduce a new vector field analogous to the
Stueckelberg field into 2-form Einstein gravity to restore the large
symmetry of its topological version.
 This new field will be useful for relating Einstein gravity to its
topological version, as the Stueckelberg field relates the massive U(1)
gauge theory to the massless one.
 Symmetries of the system with the vector field are irreducible while
those of the topological version of gravity are on-shell reducible.
 It is, however, possible to introduce an extra symmetry so that the
system has an on-shell reducibility, which is a natural extension of
the one in the topological gravity.
 We will show two different expressions for the BRST charge to develop
the BRST quantization of the system with the vector field: one of them
corresponds to the case with irreducible symmetries, and the other the
case with on-shell reducible ones including the extra symmetry.
 These BRST charges are of different rank while belonging to the same
theory. \par

\vskip 0.5cm


 The action of (Euclidean) 2-form Einstein gravity is given in terms of
2-form $\Sigma^{k}$ and SU(2) spin connection 1-form $\omega^{k}$ in
the presence of a cosmological constant $\Lambda$,
$$
S = \int {\Sigma^{k}} \wedge {R_{k}}
    -{\Lambda \over 24} \  {\Sigma^{k}} \wedge {\Sigma_{k}}
    +{\alpha \over 2} \  {\psi_{kl}}{\Sigma^{k}} \wedge {\Sigma^{l}} \ ,
                                                                \eqno(1)
$$
where $R_{k} \equiv d\omega_{k} + (\omega \times \omega)_{k}$,
$\, \psi_{kl} $ is a symmetric trace-free Lagrange multiplier field, and
$\alpha$ is a parameter.
 The SU(2) indices $i,j,k, \cdots = 1,2,3,$ in the fundamental fields
imply that they transform under the ${\it chiral}$ local-Lorentz
representation $(1,0)$ of SU(2)$\times$SU(2) [5].
 In this formulation, the metric $g_{\mu \nu}$ is defined as
\footnote{{}^{\sharp 1}}{We use the notation for the SU(2) indices,
        $F \cdot G \equiv F^i G^i$ and
        $(F \times G)^i \equiv \varepsilon_{ijk} F^j G^k$,
  where $\varepsilon_{ijk}$ is the structure constant of SU(2). }
$$
g^{1\over 2} g_{\mu \nu}
          = -{1 \over 12} \, {\epsilon^{\alpha \beta \gamma \delta}} \,
           {\Sigma_{\mu \alpha}} \cdot({\Sigma_{\beta \gamma}} \times
           {\Sigma_{\delta \nu}}) \ ,
                                  \qquad g \equiv det(g_{\mu \nu})\ .
                                                                \eqno(2)
$$
 Using this definition, we find that the action (1) is equivalent to
the usual Einstein-Hilbert action [2,6]. \par

 The topological version of the theory is obtained by simply dropping
the last term in the action (1), that is, by setting $\alpha$ = 0 [1].
 In this case, a new symmetry generated by a parameter 1-form
$\theta_1^k$ emerges in addition to diffeomorphism and the local-Lorentz
(with $\theta_0^k$) symmetries,
$$
  \delta \omega^k =D\theta_0^k +{\Lambda \over 12}\theta_1^k  \ ,
\qquad
  \delta \Sigma^k =2(\Sigma \times \theta_0 )^k + D\theta_1^k \ .
                                                                \eqno(3)
$$
 Here diffeomorphism with a vector field $\xi^\mu$ is implicitly
included in the above local-Lorentz and $\theta_1^k$- transformations as
we can see by setting
$\theta_0^k = \xi^\nu \omega_\nu^k$ and
$\theta_{1 \mu}^k = 2\xi^\nu \Sigma_{\nu \mu}^k$.
 With the appearance of the $\theta_1^k$-symmetry, the theory turns out
to be on-shell reducible in the sense that the transformation laws (3)
are invariant, modulo the equations of motion, under
$$
    \delta \theta_0^k = - {\Lambda \over 12} \epsilon_0^k \ ,
  \qquad
    \delta \theta_1^k = D \epsilon_0^k \ .                      \eqno(4)
$$
 This means that not all of the parameters in (3) are independent. \par

 {}From the view point of the topological version of gravity, one can
see that the large $\theta_1^k$-symmetry is partially broken in Einstein
gravity leaving only diffeomorphism and local-Lorentz symmetries intact
and, as a result, the modes of the gravitational wave are induced.
 The obstruction for the $\theta_1^k$-symmetry is the last term in the
action (1), which is an analogue of the mass term in the U(1) gauge
theory.
 We can restore the symmetry by introducing a Stueckelberg-type vector
field (1-form) $\eta^k$ in the last term as follows,
$$
  {\alpha \over 2} \int \, {\psi_{kl}}{\Sigma^{k}} \wedge {\Sigma^{l}}
\quad
\Rightarrow
\quad
  {\alpha \over 2} \int \, {\psi_{kl}}{\hat \Sigma}^{k} \wedge
                           {\hat \Sigma}^{l} \ ,
\quad
  {\hat \Sigma}^{k} \equiv {\Sigma^{k}}- D {\eta}^{k}+
                           {\Lambda \over 12} (\eta \times \eta)^{k} \ .
                                                                \eqno(5)
$$
 This modification makes our new system invariant under the
$\theta_1^k$-transformation in (3) with $\delta \eta^k = \theta_1^k$,
together with diffeomorphism and the local-Lorentz transformation.
 They are all independent and hence there is no reducible structure in
the system.
 But we can consider an extra symmetry generated by the transformation
with a parameter $\lambda_0$,
$$
\eqalign{
& \delta \Sigma^k
         = 2 \alpha ((\psi \cdot {\hat \Sigma}) \times \lambda_0)^k \ ,
\qquad
  \delta \eta^k = -D\lambda_0^k + {\Lambda \over 6}
   (\eta \times \lambda_0)^k \equiv -{\hat D}\lambda_0^k \ ,
\cr
& \delta \psi_{kl} = {\Lambda \over 6}(\psi_k \times \lambda_0 )^l
                    +{\Lambda \over 6}(\psi_l \times \lambda_0 )^k \ , }
                                                                \eqno(6)
$$
and $\delta \omega^k = 0$.
 Under the transformations in (4) and
$\delta \lambda_0^k = \epsilon_0^k$, an on-shell reducibility appears
also in this case, which is a natural extension of the one in the
topological gravity.
 It is easily shown that the new system is equivalent to Einstein
gravity, by choosing the gauge condition, $\eta^k=0$, for the
$\theta_1^k$-symmetry.
 The new system is, however, suitable for taking the
$\alpha \rightarrow 0$ limit because it respects the large
$\theta_1^k$-symmetry in the topological gravity.

\vskip 0.5cm


 {}From now on, we move into the Hamiltonian formalism and investigate
the BRST structure in the new system.
 The action (1) becomes in canonical form,
$$
 S = \int dt \int d^3x [
     \dot \omega_a \cdot B^a
     - \omega _0 \cdot \varphi
     - \Sigma_{a0} \cdot \phi^a
     - 2\alpha \psi_{kl} {\hat \Sigma}^k_{a0} {\hat B}^a_l] \ ,
                                                                \eqno(7)
$$
where $\omega_a^k$ and $B^a_k \, ({\hat B}^a_k) \equiv \epsilon^{abc}
\Sigma_{bc}^k \, (\epsilon^{abc} {\hat \Sigma}_{bc}^k)$
are the spatial components of the spin connection $\omega^k$ and
the 2-form $\Sigma^k \, ({\hat \Sigma^k})$.
 We solve the equation derived by varying (7) with respect to
$\psi_{kl}$ and regard it as five equations for nine $\Sigma^k_{a0}$.
 The solution is expressed by using four arbitrary variables $N^a$
and $\utilde{N}$,
$$
 \Sigma_{a0}^k =
       -{1 \over 4} \   \epsilon_{abc} [ N^b {\hat B}^c_k
       + \utilde{N}( {\hat B}^b \times {\hat B}^c )^k ]
       - {1 \over 2}{\dot \eta}_a^k
       -(\omega_0 \times \eta_a)^k
       + {1 \over 2}{\hat D}_a \eta^k_0  \ .
                                                                \eqno(8)
$$
 Substituting this result for the canonical action (7), we get five
sets of constraints:
$$
\eqalign{
 & {\hat \varphi} _k \equiv - D_aB^a_k
                     -2 (\eta_a \times {}^{\eta}\pi^a)^k \approx 0 \ ,
\qquad
   {\cal E}_k \equiv {\hat D}_a {{}^{\eta}\pi}^a_k \approx 0 \ ,
\cr
 & {\hat \phi}^a_k \equiv 2 \ ( \epsilon^{abc}R_{bc}^k
                   - {\Lambda \over 12} B^a_k ) -2 \, {{}^{\eta}\pi}^a_k
                   \approx 0 \ ,
\cr
 & H_a \equiv {1 \over 4} \  \epsilon_{abc} {\hat B}^b \cdot
              ({\hat \phi^c} + 2 \, {{}^{\eta} \pi^c}) \approx 0,
\
   {\cal H} \equiv {1 \over 4} \  \epsilon_{abc} ({\hat B}^a \times
            {\hat B}^b) \cdot ({\hat \phi^c} + 2 \, {{}^{\eta} \pi^c})
            \approx 0 \ .}
                                                                \eqno(9)
$$
 The fields ${{}^{\eta}\pi}^a_k$ are the conjugate momenta of the
spatial components of $\eta_a^k$.
The non-zero Poisson brackets among the constraints
${\hat \varphi}_k$, ${\hat \phi}^a_k$, and ${\cal E}_k$ are given by
$$
\eqalign{
  & \{ {\hat \varphi}[g_1], {\hat \varphi}[g_2] \} =
     - 2 \, {\hat \varphi}[(g_1 \times g_2)] \ ,
\qquad
    \{ {\hat \varphi}[g], {\hat \phi}^a [h_a] \} =
     - 2 \, {\hat \phi}^a [(g \times h_a)] \ ,
\cr
  & \{ {\hat \varphi}[g], {\cal E}[f] \} =
     - 2 \, {\cal E} [(g \times f)] \ ,
\qquad \quad \, \, \,
    \{ {\cal E} [f_1], {\cal E} [f_2] \} =
     - {\Lambda \over 6} \, {\cal E} [(f_1 \times f_2)] \ , }
                                                                \eqno(10)
$$
where ${\hat \varphi}[g_1] \equiv \int d^3x g_1^k {\hat \varphi}_k, \
{\hat \phi}^a [h_a] \equiv \int d^3x h_a^k {\hat \phi}^a_k $, etc.
 Next we redefine the constraint $H_a$ as
$$
 {\cal H}_a \equiv 2 H_a
                  + \omega_a \cdot {\hat \varphi}
                  - \eta_a \cdot {\cal E}
                  - {1 \over 2} \epsilon_{abc} ({\hat B}^b -B^b) \cdot
                    {\hat \phi^c} \ .
                                                                \eqno(11)
$$
 The new constraint ${\cal H}_a$ generates the spatial diffeomorphism.
 {}From (10) and (11), we can see that the Poisson algebra among
${\hat \varphi}_k$, ${\hat \phi}^a_k$, ${\cal E}_k$, and ${\cal H}_a$ is
closed.
 But the following non-zero Poisson brackets with respect to the
constraint ${\cal H}$ include fields in the structure functions and,
therefore, make the whole algebra {\it open}:
$$
\eqalign{
 & \{{\cal E}[f], {\cal H}[{\utilde N}] \} =
     - {1 \over 2} {\hat \phi}^a [\epsilon_{abc} {\utilde N}
       (f \times ({\hat B}^b \times [\phi^c
       - {\Lambda \over 12}{\hat B}^c]))]   \ ,
\cr
 & \{{\cal H}[{\utilde N}], {\cal H}[{\utilde M}] \} =
    {\cal H}_a [L^a]
    - {\hat \varphi} [L^a \omega_a ]
    + {\cal E} [L^a \eta_a ]
    + {1 \over 2} {\hat \phi}^a
      [\epsilon_{abc} L^b ({\hat B}^c - B^c )] \ ,}
                                                                \eqno(12)
$$
where $L^a \equiv {\hat B}^a \cdot {\hat B}^b ({\utilde M} \partial_b
{\utilde N} - {\utilde N} \partial_b {\utilde M})$.
 All the constraints in the system are of first class.
 Among them, ${\hat \varphi}_k$ and ${\hat \phi}^a_k$ are identified
with the generators of the local-Lorentz and $\theta_1^k$-
transformations in (3) respectively.
 The constraint ${\cal H}$ generates temporal diffeomorphism while
${\cal H}_a$ the spatial one.
 The remaining ${\cal E}_k$ corresponds to the generator of the extra
symmetry given in (6) and therefore depends on the other constraints.
 Precisely it reads
$$
  D_a {\hat \phi}^a_k - {\Lambda \over 6} {\hat \varphi}_k
  + 2 {\cal E}_k = 0 \ .                                        \eqno(13)
$$
 We can regard 2-form Einstein gravity with the $\eta_a^k$ field as
either a system with four sets of irreducible constraints, or a system
with five sets of on-shell reducible ones when including the extra
symmetry.
 In each of the cases, we give the nilpotent BRST charge by using the
formalism by Batalin, Fradkin and Vilkovisky (BFV) [7,8].
 In the following, we use the convention in [8]. \par

 Firstly we consider the irreducible case without the constraint
${\cal E}_k$.
 We have to associate the conjugate pairs of the anti-commuting
ghosts ($C^i , \bar{ {\cal P}}_i$), ($C_a^i , \bar{ {\cal P}}^a_i $),
($\xi^a, {}^{\xi}{\bar {\cal P}}_a$), and
($\utilde{\xi}, {}^{\xi}{\bar {\cal P}}$) to the constraints
$({\hat \varphi}_k, {\hat \phi}^a_k,{\cal H}_a, {\cal H})$ respectively.
 The ghost number of the ghosts $C, \xi$ (the conjugate momenta
${\bar {\cal P}}$) is 1($-$1). Using these fields, the BRST charge is
constructed as follows,
$$
 \eqalign {
 Q_{I}= \, & {\cal G}_{I} + {\cal L}_{I} +
        \int d^3x \{ C_a \cdot {\hat \phi}^a
        + {\utilde \xi} {\cal H}
        - {\utilde \xi} \partial_b {\utilde \xi} \, {\hat B}^a \cdot
          {\hat B}^b \, [{}^{\xi}{\bar {\cal P}}_a - \omega_a \cdot
          {\bar {\cal P}}
\cr
        & + \eta_a \cdot ({\Lambda \over 12} {\bar {\cal P}}
          - {1 \over 2} D_c {\bar {\cal P}}^c )
          + {1 \over 2} \epsilon_{acd} ({\hat B}^c - B^c) \cdot
          {\bar {\cal P}}^d ] \} \ ,}
                                                                \eqno(14)
$$
where
$$
 \eqalign {
 & {\cal G}_{I} \equiv \int d^3x [ C \cdot {\hat \varphi}
                - ( C \times C ) \cdot \bar{{\cal P}}
                - 2 ( C \times C_a ) \cdot \bar{{\cal P}}^a] \ ,
\cr
 & {\cal L}_{I} \equiv \int d^3x [\xi^a {\cal H}_a
           + \xi^a \partial_a \xi^b \, {}^{\xi} {\bar {\cal P}}_b
           + {\cal L}_{\xi} {\utilde \xi} \, {}^{\xi}{\bar {\cal P}}
           + {\cal L}_{\xi} C \cdot {\bar {\cal P}}
           + {\cal L}_{\xi} C_a \cdot {\bar {\cal P}}^a ] \ .}
                                                                \eqno(15)
$$
 In the Lie derivative ${\cal L}_{\xi}$ of anti-commuting fields in
(15), the ghosts $\xi^a$ are defined to be placed in front of the fields.
 The term ${\cal G}_{I}$ generates the BRST version of the local-Lorentz
transformation while ${\cal L}_{I}$ the BRST version of spatial
diffeomorphism.
 The rank of the BRST charge is one. \par

 Next we give the BRST charge in the case with on-shell reducible
constraints
$({\hat \varphi}_k, {\hat \phi}^a_k, {\cal E}_k, {\cal H}_a, {\cal H})$.
 Here we also introduce the conjugate pairs of the anti-commuting
ghosts (${}^{\eta} C^i , {}^{\eta} {\bar {\cal P}}_i$) for the
constraint ${\cal E}_k$ and the commuting `ghosts for ghosts'
(${}^{(1)}C^i , {}^{(1)}{\bar{\cal P}}_i$).
 The ghost number of these fields is
$gh({}^{\eta} C^i, {}^{\eta} {\bar {\cal P}}_i, {}^{(1)}C^i,
{}^{(1)}{\bar {\cal P}}_i)$ $=(1,-1,2,-2)$.
 In this case, the BRST charge is given by
$$
\eqalign{
 & Q_{II} = {\cal G}_{II} + {\cal L}_{II}
      + \int d^3x \{ C_a \cdot {\hat \phi}^a
      + {}^{\eta}C \cdot {\cal E}
      + {\utilde \xi} {\cal H}
      - {\Lambda \over 12} ({}^{\eta} C \times {}^{\eta} C)
        \cdot {}^{\eta}{\bar {\cal P}}
\cr
    & - {1 \over 2} \epsilon_{abc} {\utilde \xi}[({\cal B}^a \times
        (\phi^b-{\Lambda \over 12} {\cal B}^b))
        + {1 \over 2} ( \phi^a \times ({}^{\eta} C \times
        {\bar {\cal P}}^b))] \cdot ({}^{\eta}C \times {\bar {\cal P}}^c)
\cr
    & - {\utilde \xi} \partial_b {\utilde \xi} \, {\cal B}^a \cdot
        {\cal B}^b [{}^{\xi}{\bar {\cal P}}_a
        - \omega_a \cdot {\bar {\cal P}}
        + \eta_a \cdot {}^{\eta}{\bar {\cal P}}
        + {1 \over 2} \epsilon_{acd} ({\hat B}^c - B^c) \cdot
        {\bar {\cal P}}^d
        - i C_a \cdot {}^{(1)}{\bar {\cal P}}]
\cr
    & + i \, {}^{(1)}C \cdot ( D_a \bar{{\cal P}}^a
        - {\Lambda \over 6} {\bar {\cal P}}
        + 2 \, {}^{\eta} {\bar {\cal P}} ) \} \ ,}
                                                                \eqno(16)
$$
where
$$
\eqalign {
 & {\cal G}_{II} \equiv {\cal G}_{I}
  - \int d^3x [2 (C \times {}^{\eta} C) \cdot {}^{\eta}{\bar {\cal P}}
             + 2 (C \times {}^{(1)}C) \cdot {}^{(1)}{\bar{\cal P}} ] \ ,
\cr
 & {\cal L}_{II} \equiv {\cal L}_{I}
  + \int d^3x[{\cal L}_{\xi} {}^{\eta} C \cdot {}^{\eta}{\bar {\cal P}}
         + {\cal L}_{\xi} {}^{(1)}C \cdot {}^{(1)}{\bar {\cal P}} ] \ ,}
                                                                \eqno(17)
$$
and ${\cal B}^a_k \equiv {\hat B}^a_k - ({}^{\eta} C \times
{\bar {\cal P}}^a )_k $.
 The rank of this $Q_{II}$ is three.
 Both the BRST charges in (14) and (16) satisfy the nilpotency condition
$\{Q,\ Q \}= 0 $ and are real, anti-commuting and $gh(Q) = 1$.
 The reducibility of constraints has made the expression for $Q_{II}$
in (16) somewhat complicated.
 But we can choose another set of constraints with off-shell
reducibility in our same system so that the corresponding BRST charge
has the rank one [9]. \par

 To summarize, we have introduced the vector field $\eta^k$ into 2-form
Einstein gravity in four dimensions to restore the large symmetry of its
topological version.
 In the new system, we have obtained two different expressions for the
BRST charge: one of them ($Q_{I}$) for irreducible symmetries and the
other ($Q_{II}$) for on-shell reducible ones.
 The latter $Q_{II}$ will be useful for our approach to gravity from its
topological version in quantum theory [9]. \par

\vskip 0.3cm

 We are grateful to Dr. S. Yahikozawa for important suggestions and
discussions.
 We would like to thank Professor A. Hosoya for careful reading of the
manuscript.

\vfill

\eject



\REF\SEELNU{H.Y. Lee, A. Nakamichi, and T. Ueno, to be published in
Phys.$\>$Rev.$\>$D47 (1993).}

\REF\SEECDJM{R. Capovilla, J. Dell, T. Jacobson and L. Mason,  \nextline
Class.$\>$Quantum$\>$Grav.$\>$8 (1991) 41.}

\REF\SEEA{A. Ashtekar, Phys.$\>$Rev.$\>$D36 (1987) 1587; {\it `New
          Perspectives in Canonical Gravity'} (Bibliopolous, Naples,
          Italy, 1988).}

\REF\SEES{E.G. Stueckelberg, Helv.$\>$Phys.$\>$Acta$\>$30 (1957) 209.}

\REF\SEEPR{R. Penrose and W. Rindler, {\it `Spinors and Space-time'}
Vols. I, II (Cambridge University Press, Cambridge, 1984).}

\REF\SEEJS{T. Jacobson and L. Smolin, Class.$\>$Quantum$\>$Grav.$\>$5
(1988) 583.}

\REF\SEEFVBV{E.S. Fradkin and G.A. Vilkovisky, Phys.$\>$Lett.$\>$B55
(1975) 224;
 \nextline   I.A. Batalin and G.A. Vilkovisky, Phys.$\>$Lett.$\>$B69
(1977) 309.}

\REF\SEEH{M. Henneaux, Phys.$\>$Rep.$\>$26 (1985) 1.}

\REF\SEENU{A. Nakamichi and T. Ueno, in preparation.}

\refout

\end